\documentclass[]{elsart5p}
\usepackage[]{graphicx}
\usepackage{amssymb}

\begin{document}

\begin{frontmatter}

\title{Generalized exponential function and discrete growth models}

\author[alexandre1]{Alexandre Souto Martinez}
\address[alexandre1]{Faculdade de Filosofia, Ci\^encias e Letras de Ribeir\~ao Preto,\\ 
Universidade de S\~ao Paulo, and \\
National Institute of Science and Technology for Complex Systems \\
Avenida Bandeirantes, 3900\\
14040-901, Ribeir\~ao Preto, S\~ao Paulo, Brazil.}
\ead{asmartinez@ffclrp.usp.br}

\author[rodrigo]{Rodrigo Silva Gonz\'alez}
\address[rodrigo]{Faculdade de Filosofia, Ci\^encias e Letras de Ribeir\~ao Preto,\\ 
Universidade de S\~ao Paulo,\\
Avenida Bandeirantes, 3900\\
14040-901, Ribeir\~ao Preto, S\~ao Paulo, Brazil.}
\ead{caminhos\_rsg@yahoo.com.br}

\author[aquino]{Aquino Lauri Esp\'{\i}ndola}
\address[aquino]{Faculdade de Filosofia, Ci\^encias e Letras de Ribeir\~ao Preto,\\ 
Universidade de S\~ao Paulo,\\
Avenida Bandeirantes, 3900\\
14040-901, Ribeir\~ao Preto, S\~ao Paulo, Brazil.}
\ead{aquinoespindola@usp.br}

\journal{Physica A}

\begin{abstract}
Here we show that a particular one-parameter generalization of the exponential function is suitable to unify most of the popular one-species discrete population dynamics models into a simple formula.
A physical interpretation is given to this new introduced parameter in the context of the continuous Richards model, which remains valid for the discrete case. 
From the discretization of the continuous Richards' model (generalization of the Gompertz and Verhuslt models), one obtains a generalized logistic map and we briefly study its properties. 
Notice, however that the physical interpretation for the introduced parameter persists valid for the discrete case.
Next, we generalize the (scramble competition) $\theta$-Ricker discrete model and analytically calculate the fixed points as well as their stability.
In contrast to previous generalizations, from the generalized $\theta$-Ricker model one is able to retrieve either scramble or contest models. 
\end{abstract}

\end{frontmatter}

\section{Introduction}

Recently, the generalizations of the logarithmic and exponential functions have attracted the attention of resear\-chers. 
One-parameter logarithmic and exponential functions have been proposed in the context of non-extensive statistical mechanics~\cite{tsallis_1988,tsallis_qm,nivanen_2003,borges_2004,kalogeropoulos_2005}, relativistic statistical mechanics~\cite{kaniadakis_2001,PhysRevE.66.056125} and quantum group theory~\cite{abe_1997}. 
Two and three-parameter generalization of these functions have also been proposed~\cite{kaniadakis:046128,kaniadakis:036108,veit:2007}. 
These generalizations are in current use in a wide range of disciplines since they permit the generalization of special functions: hyperbolic and trigonometric~\cite{borges_1998}, Gaussian/Cauchy probability distribution function~\cite{tsallis_levy} etc.   
Al\-so, they permit the description of several complex systems~\cite{tsallis_stariolo:1996,albuquerque:2000,cajueiro_2006,cajueiro_2007,anteneodo:1:2002,holanda:2004}, for instance in generalizing the stretched exponential function~\cite{martinez:2008c}. 

As mentioned above, the one-parameter generalizations of the logarithm and exponential functions are not univoquous. 
The $\tilde q$-logarithm function $\ln_{\tilde q}(x)$ is defined as the value of the area underneath the non-symmetric hyperbole,  $f_{\tilde q}(t)=1/t^{1-\tilde q}$, in the interval $t \in [1,x]$ \cite{tiago}:
\begin{eqnarray}
\ln_{\tilde q}(x) & = & \int_1^x \frac{dt}{t^{1-{\tilde q}}}=\lim_{{\tilde q}^\prime \to {\tilde q}}\frac{x^{\tilde q^\prime}-1}{{\tilde q}^\prime} 
\; .
\label{eq:gen_log}
\end{eqnarray}
This function is {\it not} the ordinary logarithmic function in the basis $\tilde q$, namely $[\log_{\tilde q}(x)]$, but a generalization of the natural logarithmic function definition, which is recovered for $\tilde q=0$. 
The area is negative for $0<x<1$, it vanishes for $x=1$ and it is positive for $x>1$, independently of the $\tilde q$ values. 

Given the area $x$ underneath the curve $f_{\tilde q}(t)$, for $t\in [0,y]$, the upper limit $y$ is the generalized $\tilde q$-exponential function: $y=e_{\tilde q}(x)$. This is the inverse function of the $\tilde q$-logarithmic $e_{\tilde q}[\ln_{\tilde q}(x)]=x=\ln_{\tilde q}[e_{\tilde q}(x)]$ and it is given by:
\begin{equation}
 e_{\tilde q}(x)  = 
 \left\{
 \begin{array}{ll}
0                                                                                            & \; \mbox{for} \;  \tilde{q} x < -1 \\
 \lim_{{\tilde q}^\prime \to {\tilde q}}(1+{\tilde q}^\prime x)^{1/{\tilde q}^\prime}        &  \; \mbox{for} \;  \tilde{q} x \ge -1 
\end{array}
\right. \; .
\label{eq:limite}
\end{equation}
This is a non-negative function $e_{\tilde q}(x) \geq 0$, with $e_{\tilde q}(0)=1$, for any $\tilde q$.
For $\tilde q \to \pm \infty$, one has that $e_{- \infty}(x)=1$, for $x \le 0$ and $e_{\infty}(x)=1$, for $x \ge 0$.  
Notice that letting $x=1$ one has generalized the Euler's number:
\begin{equation}
 e_{\tilde q} (1) = (1+\tilde q)^{1/\tilde q}.
\label{eq:eqtilde}
\end{equation}

Instead of using the standard entropic index $q$ in Eqs.~(\ref{eq:gen_log}) and (\ref{eq:eqtilde}), we have adopted the notation $\tilde{q} = 1 - q$. 
The latter notation permits us to write simple relations as: $\ln_{\tilde{q}}(x) = - \ln_{-\tilde{q}}(x)$ or $e_{-\tilde{q}}(-x) = 1/e_{\tilde{q}}(x)$, bringing the inversion point around $\tilde{q} = 0$.
These relations lead to simpler expressions in population dynamics problems~\cite{martinez:2008b} and the generalized stretched exponential function~\cite{martinez:2008c} contexts. 
Also, they simplify the generalized sum and product operators~\cite{tiago}, where a link to the aritmethical and geometrical averages of the generalized functions is established.

This logarithm generalization, as shown in Ref.~\cite[p. 83]{montroll_west}, is the one of non-extensive statistical mechanics~\cite{tsallis_qm}. 
It turns out to be precisely the form proposed by Montroll and Badger~\cite{badger_1974} to unify the Verhulst ($\tilde q = 1$) and Gompertz ($\tilde q = 0$) one-species population dynamics model. 
The $\tilde{q}$-logarithm leads exactly to the Richards' growth model~\cite{richards_1959,martinez:2008b}:
\begin{equation}
 \frac{d \ln p(t)}{dt} = -\kappa \ln_{\tilde q}p(t),
 \label{eq:richard_model}
\end{equation}
where $p(t)=N(t)/N_\infty$, $N(t)$ is the population size at time $t$, $N_\infty$ is the carrying capacity and $\kappa$ is the intrinsic growth rate. The solution of Eq.~(\ref{eq:richard_model}) is the {\it $\tilde q$-generalized logistic} equation  $p(t) = 1/{e_{\tilde q}[\ln_{\tilde q}(p_0^{-1})e^{-\kappa t}]}  = e_{-{\tilde q}}[-\ln_{\tilde q}(p_0^{-1})e^{-\kappa t}] =  e_{-{\tilde q}}[\ln_{-\tilde q}(p_0)e^{-\kappa t}]$.

The competition among cells drive to replicate and inhibitory interactions, that are modeled by long range interaction among these cells.
These interactions furnish an interesting microscopic mechanism to obtain Richards' model~\cite{idiart_2002,idiart_2002-2}. 
The long range interaction is dependent on the distance $r$ between two cells as a power law $r^{\gamma}$.
These cells have a fractal structure characterized by a fractal dimension $D_f$. 

Here we call the attention to Eq.~(7) of Ref.~\cite{idiart_2002}, namely $\dot n(t)= n(t)\{\left<G\right> - JI[n(t)]\}$, where $I(n(t)) = \omega\left\{[D_f n(t)/\omega]^{1-\gamma/D_f}-1\right\}/[D_f(1-\gamma/D_f)]$.
Here, $\omega$ is a constant related to geometry of the problem, $\left< G \right>$  is the mean intrinsic replication rate of the cells and $J$ is the interaction factor. Using Eq.~(\ref{eq:gen_log}), one can rewrite it simply as: $\mbox{d}\ln n(t)/\mbox{d}t = \langle G \rangle / n(t)- J \omega \ln_{\tilde{q}}[D_f n(t)/\omega]/{D_f}$. 
Calling, $p = D_f n/\omega$, $\kappa = J \omega/D_f$ and $\tilde{q} = 1 - \gamma/D_f$, this equation is the Richard's model [Eq.~(\ref{eq:richard_model})] with an effort rate $\langle G \rangle / n(t)$.   
In this context the parameter $\tilde{q}$ acquires a physical meaning related to the interaction range $\gamma$ and fractal dimension of the cellular structure $D_f$.
If the interaction does not depend on the distance, $\gamma=0$, and it implies that $\tilde q=1$.
This physical interpretation of $\tilde{q}$ has only been possible due to Richards' model underlying microscopic description.

Introduced by Nicholson in 1954~\cite{hassell_1975}, scramble and contest are types of intraspecific competition models that differ between themselves in the way that limited resources are shared among individuals. 
In scramble competition, the resource is equally shared among the individuals of the population as long as it is available. 
In this case, there is a critical population size $N_c$, above which, the amount of resource is not enough to assure population survival.
In the contest competition, stronger individuals get the amount of resources they need to survive. 
If there is enough resources to all individuals, population grows, otherwise, only the strongest individuals survive (strong hierarchy), and the population maintains itself stable with size $N_\infty$.

From experimental data, it is known that other than the important parameter $\kappa$ (and sometimes $N_\infty$), additional parameters in more complex models are needed to adjust the model to the given population. 
One of the most general discrete model is the $\theta$-Ricker model~\cite{bellows_1981,berryman_1999}. 
This model describes well scramble competition models but it is unable to put into a unique formulation the contest competition models such as Hassel model~\cite{hassell_1975}, Beverton-Holt model \cite{beverton_holt} and Maynard-Smith-Slatkin model~\cite{maynard-smith_1973}. 

Our main purpose is to show that Eq. (\ref{eq:limite}) is suitable to unify most of the known discrete growth models into a simple formula. 
This is done in the following way. 
In Sec.~\ref{sec:loquistic}, we show that the Richards' model [Eq. (\ref{eq:richard_model})], which has an underlying microscopic model, has a physical interpretation to the parameter $\tilde{q}$, and its discretization leads to a generalized logistic map.  
We briefly study the properties of this map and show that some features of it (fixed points, cycles etc.) are given in terms of the $\tilde q$-exponential function.  
Curiously, the map attractor can be suitably written in terms of $\tilde{q}$-exponentials, even in the logistic case. 
In Sec.~\ref{sec:generalized_theta_ricker}, using the $\tilde q$-exponential function, we generalize the $\theta$-Ricker model and analytically calculate the model fixed points, as well as their stability. 
In Sec.~\ref{sec:generalizedskellam}, we consider the generalized Skellam model. 
These generalizations allow us to recover most of the well-known scramble/contest competition models. 
Final remarks are presented in Sec~\ref{sec:conclusion}.

\section{Discretization of the Richards' model}
\label{sec:loquistic}

To discretize Eq.~(\ref{eq:richard_model}), call $(p_{i+1}-p_i)/\Delta t =  -kp_i(p_{i}^{\tilde q}-1)/{\tilde q},~\rho_{\tilde q}^{\prime}= 1 + k\Delta t/{\tilde q}$ and $x_i= p_i[(\rho_{\tilde q}-1)/\rho_{\tilde q}]^{\tilde q}$, which leads to:
\begin{equation}
 x_{i+1}=\rho_{\tilde q}^{\prime}x_i(1-x_i^{\tilde q})=-\rho_{\tilde q}x_i\ln_{\tilde q}(x_i)\;, 
\label{eq:loquistic}
\end{equation}
where $\rho_{\tilde q}={\tilde q}\rho_{\tilde q}^{\prime}$.
We notice that $\tilde q$ keeps its physical interpretation of the continuous model.

In Eq.~(\ref{eq:loquistic}), if $\tilde{q} = 1$ and $\rho_1 = \rho_1^{\prime} = 4 a$, with $a \in [0,1]$, one obtains the \emph{logistic map}, $x_{i+1} = 4 a x_i (1 - x_i)$, which is the classical example of a {\it dynamic system} obtained from the discretization of the Verhulst model. 
Although simple, this map presents a extremely rich behavior, universal period duplication, chaos etc.~\cite{may_1976}.  

Let us digress considering the Feigenbaum's map~\cite{feigenbaum_1979}: $y_{i+1} = 1 - \mu y_i^{\tilde{q} + 1}$, with $\tilde{q} > 0$, $0 < \mu \le 2$ and $-1 \le y_{i} \le 1$. 
Firstly, let us consider the particular case $\tilde{q} = 1$. 
If one writes $y_i = \tilde{y}_i - b$, with $b$ being a constant, then: $\tilde{y}_{i+1} = - \mu b^2 + b + 1 + 2 \mu b  \tilde{y}_i[1 - \tilde{y}_i/(2b)]$. 
Imposing that $ - \mu b^2 + b + 1 = 0$  leads to $b_{\pm} = (1 \mp \sqrt{1 + 4 \mu})/(2 \mu)$ and calling $x_i = \tilde{y}_i/(2b)$, one obtains the logistic map with $\rho_1 = \rho'_1 = 4a = 1 + \sqrt{1 + 4 \mu}$, so that $0 < \rho_1 \le 4$. 
One can easily relate the control parameter of these two maps, making the maps equivalent. 
For arbitrary values of $\tilde{q}$, there is not a general closed analytical form to expand $|\tilde{y}_i - b|^{\tilde{q} + 1}$ and one cannot simply transform the control parameters of Eq.~(\ref{eq:loquistic}) to the Feigenbaum's map.
Here, in general, these two maps are not equivalent. 
It would be then interesting, to study the sensitivity of Eq.~(\ref{eq:loquistic}) with respect to initial conditions as it has been extensively studied in the Feigenbaum's map~\cite{lyra_1997,lyra_1998,lyra_1999,lyra_2000}.

Returning to Eq.~(\ref{eq:loquistic}),  in the domain $0\leq x \leq 1$, $f(x_i)=-\rho_{\tilde q}x_i\ln_{\tilde q}(x_i)\geq 0$ (non-negative), for $\rho_{\tilde q} > 0$. 
Since $e_{\tilde{q}}(x)$ is real only for $\tilde{q} x > -1$, $\tilde{f}$ is real only for $\tilde{q} > -1$. 
The maximum value of the function is 
\begin{equation}
\tilde{f} = f(\tilde{x}) = \frac{\rho_{\tilde q}}{e_1({\tilde q})e_{\tilde q}(1)} \; , 
\end{equation}
which occurs at 
\begin{equation}
\tilde x = \frac{1}{e_{\tilde q}(1)} \; ,
\end{equation}
i. e., the inverse of the generalized Euler's number $e_{\tilde q}(1)$ [Eq.~(\ref{eq:eqtilde})]. 
For the generalized logistic map, $0 \le x \le 1$, so that $0 \le \tilde{f} \le 1$, it leads to the following domain for the control parameter $0 \le \rho_{\tilde{q}} \le \rho_{max}$:
\begin{equation}
\rho_{max} = e_{\tilde{q}}(1)  e_1(\tilde{q})  = (1 + \tilde{q})^{1+1/\tilde{q}} \; .
\label{eq:rho_max}
\end{equation}

The map fixed points $[x^*=f(x^*)]$ are 
\begin{eqnarray}
x_1^* & = & 0, \label{eq:fp1} \\ 
x_2^* & = & e_{\tilde q}(-1/{\rho_{\tilde q}}) \label{eq:fp2} \; .
\end{eqnarray}
The fixed point $x_1^*$ is stable for $0 \leq \rho_{\tilde q} < {\tilde q}$ and $x_2^*$ is stable for ${\tilde q} \leq \rho_{\tilde q} < \rho_{pd}$, where 
\begin{equation}
\rho_{pd} = \tilde{q} + 2 \; .
\label{eq:rho_pd}
\end{equation}
Notice the presence of the $\tilde{q}$-exponentials in the description of the attractors, even for the logistic map $\tilde{q} = 1$. 

\begin{figure}[hbt]
\begin{center}
\includegraphics[width = \columnwidth]{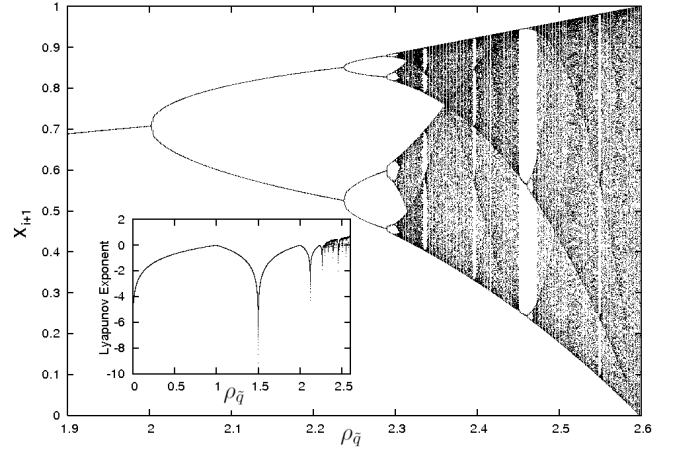}
\caption{Bifurcation diagram of Eq.~(\ref{eq:loquistic}) for $\tilde{q} = 2$, where we see that the period doubling start at $\rho = (\tilde{q} + 2)/\tilde{q} = 2$ [Eq.~(\ref{eq:rho_pd})] and the chaotic phase finishes at $\rho_{max} = e_{2}(1) e_1(2)/\tilde{q} = \sqrt{27}/2 \approx 2.6$ [Eq.~(\ref{eq:rho_max})]. {\bf Inset:} Lyapunov exponents as function of $\rho_{\tilde q}$ for $\tilde q = 2$.}
\label{fig1}
\end{center}
\end{figure}

The generalized logistic map also presents the rich behavior of the logistic map as depicted by the bifurcation diagram of Fig.~\ref{fig1}.
The inset of Fig.~\ref{fig1} displays the Lyapunov exponents as function of the central parameter $\rho_{\tilde q}$.

\begin{figure}[hbt]
\begin{center}
\includegraphics[ width = \columnwidth]{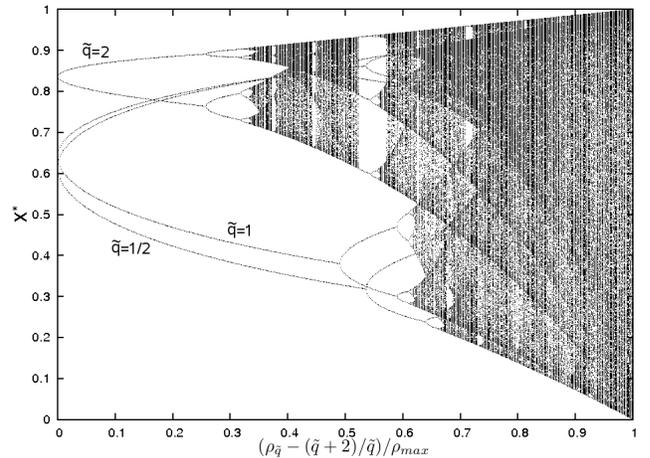}
\caption{Bifurcation diagram of Eq.~(\ref{eq:loquistic}) for $\tilde{q} = 1/2$, $\tilde{q} = 1$ (logistic) and $\tilde{q} = 2$. The fixed points are given by Eqs.~(\ref{eq:fp1}) and~(\ref{eq:fp2}).}
\label{fig2}
\end{center}
\end{figure}

In Fig.~\ref{fig2} we have scaled the axis to $\rho_{\tilde{q}}[- (\tilde{q}+2)/\tilde{q}]/ (\rho_{max}\tilde{q})$, where $\rho_{max}$ is given by Eq.~(\ref{eq:rho_max}) and we plotted the bifurcation diagram for $\tilde{q} = 1/10, 1$ and $10$. 
We see that the diagrams display the same structure but each one has its own scaling parameters. 
The role of increasing $\tilde{q}$ is to lift the bifurcation diagram to relatively anticipating the chaotic phase. 
The period doubling region start at $x_2^{*}(\tilde{q}) = e_{\tilde{q}}[-1/(\tilde{q} + 2) = [1 - 1/(1 + 2/\tilde{q})]^{1/\tilde{q}}$, so that for $x_2^{*}(1/10) = (20/21)^{10} \approx 0.61$, $x_2^{*}(1) = 2/3 \approx 0.67$ and $x_2^{*}(10) = (1/6)^{1/10} \approx 0.84$. 

\begin{figure}[hbt]
\begin{center}
\includegraphics[width = \columnwidth]{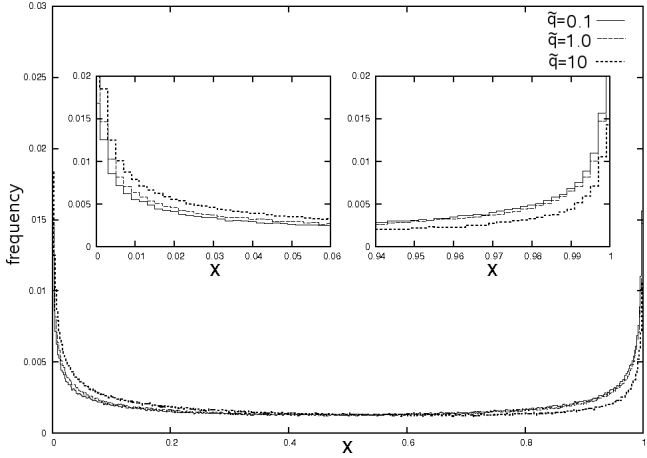}
\caption{Histograms of $x$ for $\rho_{\tilde{q}} = e_{\tilde{q}}(1)  e_1(\tilde{q})$ in Eq.~(\ref{eq:loquistic}) and $\tilde{q} = 1/10$, $\tilde{q} = 1$ (logistic) and $\tilde{q} = 10$. There is not a drastic change with respect to the logistic map. Only on the corners one is able to see the difference as shown by the insets.}
\label{fig3}
\end{center}
\end{figure}

When $\rho_{\tilde{q}} = e_{\tilde{q}}(1)  e_1(\tilde{q})$, then $x_i \in (0,1)$. 
In Fig.~\ref{fig3} we show the histograms of the distribution of the variable $x_i$.
We see that as $\tilde{q}$ increases, the histograms have the same shape as the logistic histogram has, but it is crooked in the counter clock sense around $x=1/2$.  

\section{The generalized $\theta$-Ricker model}
\label{sec:generalized_theta_ricker}

The {\it $\theta$-Ricker} model~\cite{bellows_1981,berryman_1999} is given by:
\begin{equation}
 x_{i+1} = x_i e^{r[1-(x_i/\kappa)^\theta]}, 
 \label{eq:theta_ricker}
\end{equation}
where $\theta > 0$. 

Notice that $\tilde x= r^{1/\theta} x/\kappa $ is the relevant variable, where $\kappa_1 = e^r > 0$. 
In this way Eq.~(\ref{eq:theta_ricker}) can be simply written as $\tilde x_{i+1}=k_1\tilde x_ie^{-\tilde x_i^\theta}$.
For $\theta=1$, one finds the standard {\it Ricker} model~\cite{ricker_1954}. 
For arbitrary $\theta$, expanding the exponential to the first order one obtains the generalized logistic map [Eq.~(\ref{eq:loquistic})] which becomes the logistic map, for $\theta=1$. 
The $\theta$-Ricker, Ricker and quadratic models are all scramble competion models.

If one switches the exponential function for the ${\tilde q}$-gene\-ralized exponential in Eq.~(\ref{eq:theta_ricker}), one gets the {\it generalized $\theta$-Ricker model}:
\begin{equation}
x_{i+1} = \kappa_1 x_i \; e_{-{\tilde q}} \left[ -r \, \left( \frac{x_i}{\kappa} \right)^\theta \right] 
        = \frac{\kappa_1 x_i}{\left[1 + \tilde{q} r \left( \frac{x_i}{\kappa}\right)^{\theta} \right]^{1/\tilde{q}}} \; .
\label{eq:generalized_theta_ricker_model}
\end{equation}
To obtain standard notation, write $c = 1/{\tilde q}$ and $k_2=r/(kc)$, so that $x_{i+1} = k_1x_i/(1+k_2x_i)^c$ \cite{brannstrom_2005}.

The generalized model with $\theta=1$, leads to the {\it Hassel} model~\cite{hassell_1975}, which can be a scramble or contest competition model. 
One well-known contest competition model is the {\it Beverton-Holt} model~\cite{beverton_holt},  which is obtained taking ${\tilde q}=c=1$. 
For ${\tilde q}=0$, one recovers the Ricker model and for ${\tilde q}=-1$, one recovers the logistic model.

It is interesting to mention that the Beverton-Holt model~\cite{beverton_holt}  is one of the few models that have the time evolution explicitly written: $\tilde{x}_{i} = \kappa_1^{i} \tilde{x}_0/[1 + (1 - \kappa_1^{i})\tilde{x}_0/(1 - \kappa_1)]$. 
From this equation, one sees that $x_{i (\gg 1)} = 0$ ,for $\kappa_1 \le 1$ and $x_{i (\gg 1)} = \kappa_1 - 1$ for $\kappa_1 \ge 1$.

Using arbitrary values of $\theta$ in Eq.~(\ref{eq:generalized_theta_ricker_model}), for ${\tilde q}=0$ one recovers the $\theta$-Ricker model, and for ${\tilde q}=1$, the {\it Maynard-Smith-Slatkin} model~\cite{maynard-smith_1973} is recovered.
The latter is a scramble/contest competition model. 
For ${\tilde q}=-1$, one recovers the generalized logistic map.
The trivial linear model is retrieved for $\tilde q \to -\infty$.

\begin{figure}[ht]
\begin{center}
\includegraphics[angle=-90, width=.9\columnwidth]{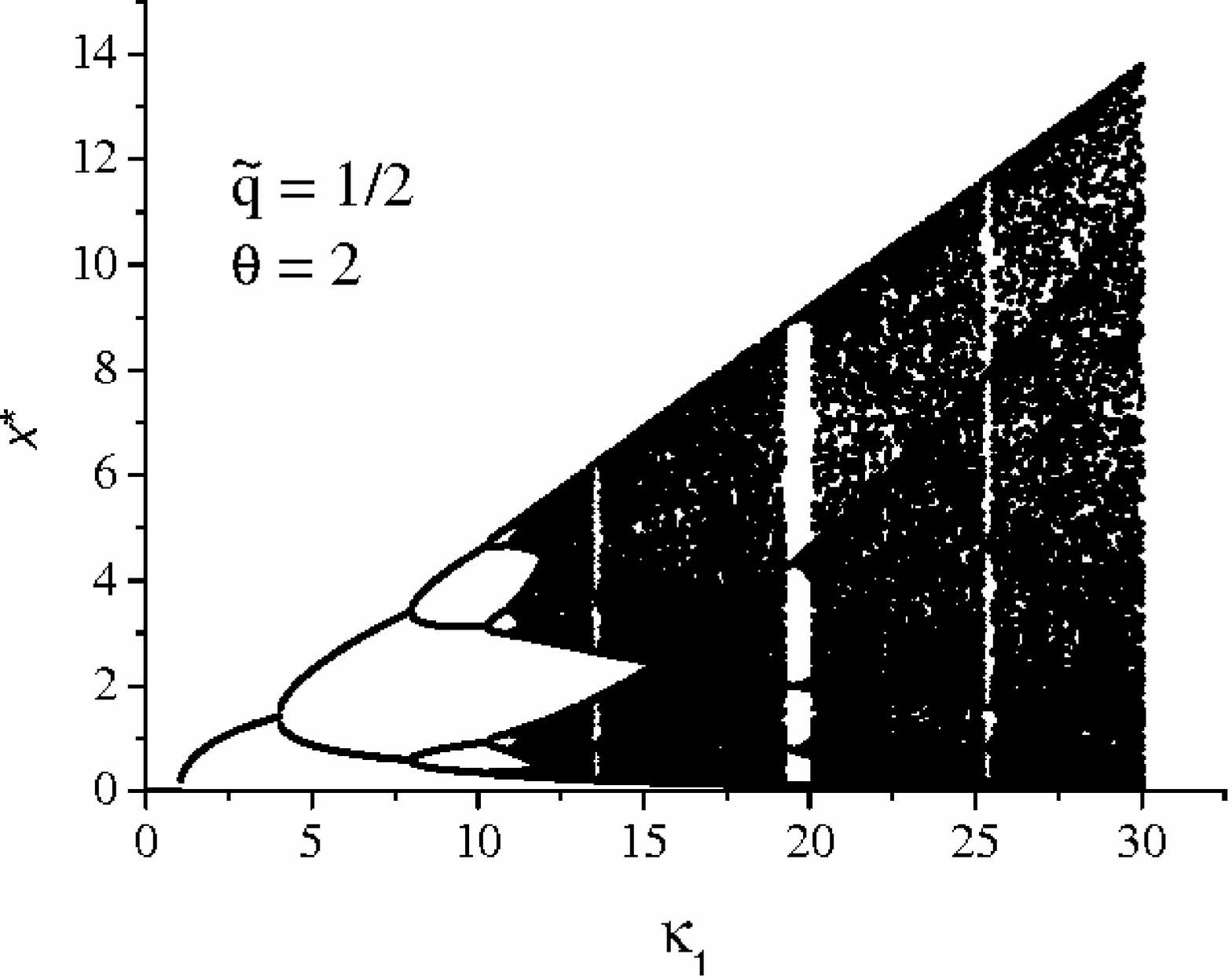}
{\bf (a)}
\includegraphics[angle=-90, width=.9\columnwidth]{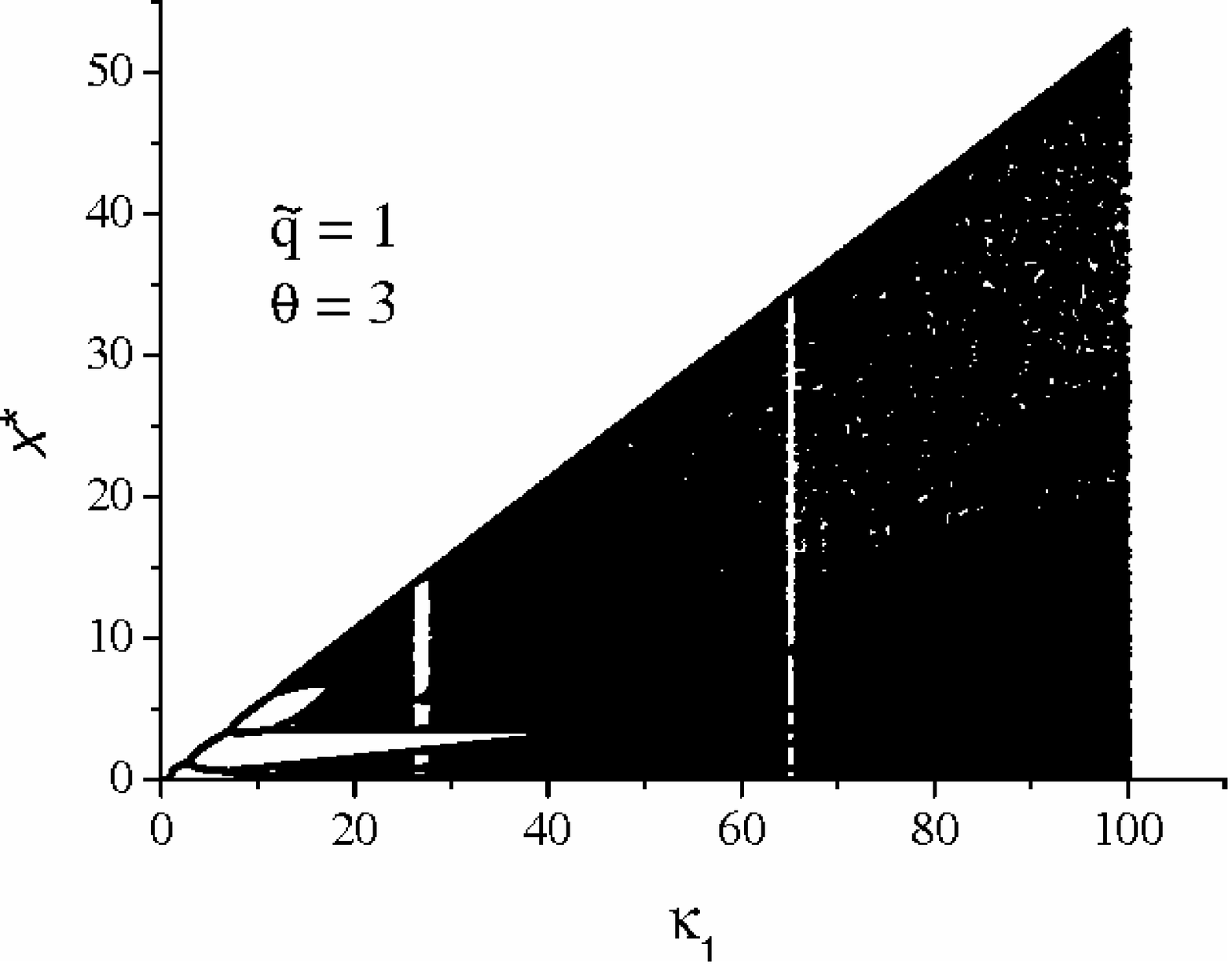}
{\bf (b)}
\includegraphics[angle=-90, width=.9\columnwidth]{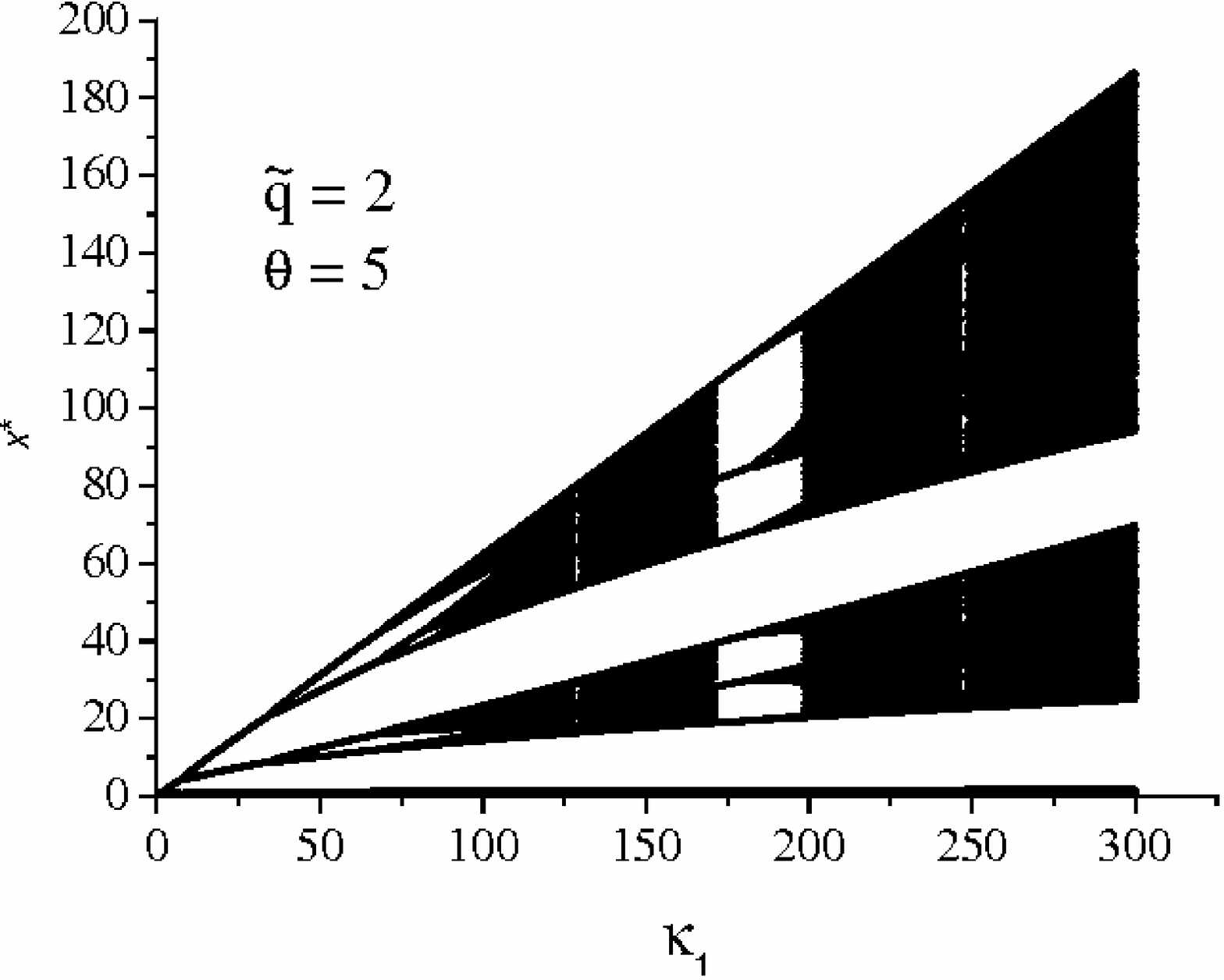}
{\bf (c)}
\caption{Typical bifurcation diagrams of Eq.~(\ref{eq:final}). 
{\bf (a)} $\tilde{q} = 1/2$ and $\theta = 2$, {\bf (a)} $\tilde{q} = 1$ and $\theta = 3$ and {\bf (a)} $\tilde{q} = 2$ and $\theta = 5$.}
\label{figbdtrm}
\end{center}
\end{figure}

In terms of the relevant variable $\tilde{x}$, Eq.~(\ref{eq:generalized_theta_ricker_model}) is rewritten as:
\begin{equation}
\tilde{x}_{i+1} = \kappa_1 \tilde{x}_i e_{-{\tilde q}}(- \tilde{x}_i^\theta) \; ,
\label{eq:final}
\end{equation}
where  $\tilde{x}_i \ge 0$  and we stress that the important parameters are $\kappa_1 > 0$, $\tilde{q}$ and $\theta > 0$. 
Eq.~(\ref{eq:final}) is suitable for data analysis and the most usual known discrete growth models are recovered with the judicious choice of the $\tilde q$ and $\theta$ parameters as it shown in Table~\ref{tabela}.
Some typical bifurcation diagrams of Eq.~(\ref{eq:final}) are displayed in Fig.~\ref{figbdtrm}.

\begin{table}[hbt]
\begin{center}
\begin{tabular}{l|c|c|l}
\hline
\hline
Model                & $\tilde q$              & $\theta$   &  competition  \\
                          &                             &                 &   type \\
\hline
\hline
Linear                & $-\infty$               &~$>0~$        &                \\
\hline
Logistic             &  $-1$                    & $1$           &~s             \\
\hline
Generalized Logistic &   $-1$          & $>0$          &~s  \\
\hline
Ricker               &  $0$                    & $1$             &~s                \\
\hline 
$\theta$-Ricker &  $0$                    & $>0$          &~s                    \\
\hline
Hassel              &   $*$                    &  1               &~s ($\tilde{q} < 1/2$) or c ($\tilde{q} \ge 1/2$) \\
\hline
Maynard-Smith-Slatkin &   $1$       & $ > 0$        &~s ($\theta > 2$ ) or c ($\theta \le 2$) \\
\hline
Beverton-Holt   &  $1$                  &     $1$        &~c\\
\hline
\hline
\end{tabular}
\caption{Summary of the parameters to obtain discrete growth models from Eq.~(\ref{eq:final}). In the competition type column, \emph{s} and \emph{c} stand for scramble and contest models, respectively. The symbol $*$ stands for arbitrary values.}
\label{tabela}
\end{center}
\end{table}

Now, let us obtain some analytical results for the map of Eq.~(\ref{eq:final}), which we write as $\tilde{x}_{i+1} = f_{gtr}(\tilde{x}_i)$, with 
\begin{equation}
f_{gtr}(\tilde{x}) = \kappa_1 \tilde{x} e_{-{\tilde q}}(- \tilde{x}^\theta) = \frac{\kappa_1  \tilde{x}}{(1 + \tilde{q} \tilde{x}^{\theta})^{1/\tilde{q}}}\; .
\label{eq:mapa_generalizado}
\end{equation} 
The $\tilde{x}$ domain is unbounded ($\tilde{x} \ge 0$), for $\tilde{q}  \ge 0$. 
However, for $\tilde{q}  < 0$, $f_{gtr}(\tilde{x}) = \kappa_1  \tilde{x} (1 - |\tilde{q}| \tilde{x}^{\theta})^{1/|\tilde{q}|}$ and the $\tilde{x}$-domain  is bounded to the interval: $0 \le \tilde{x}  \le \tilde{x}_m$, with 
\begin{equation}
\tilde{x}_m = \frac{1}{(- \tilde{q})^{1/\theta}} \; ,
\label{eq:xm}
\end{equation} 
so that for $|\tilde{q}| < 1$, $\tilde{x}_m > 1$; 
for $|\tilde{q}| = 1$,  with $\tilde{x}_m = 1$ (for $\tilde{q} = -1$, it is the generalized logistic case [Eq.~(\ref{eq:loquistic}) in $\theta$ instead of $\tilde{q}$]  and  for $\tilde{q} = 1$, the Maynard-Smith-Slatkin model) and  
for $|\tilde{q}| > 1$, $\tilde{x}_m < 1$.

The derivative of $f_{gtr}$ with respect to $\tilde{x}$ is: $f_{gtr}'(\tilde{x}) = \kappa_1 [1 + (\tilde{q} - \theta)\tilde{x}^{\theta}]/(1 + \tilde{q} \tilde{x}^{\theta})^{1+1/\tilde{q}}$. 
Imposing $f_{gtr}'(\tilde{x}_{max}) = 0$, one obtains the maximum of Eq.~(\ref{eq:mapa_generalizado}),  
\begin{equation}
\tilde{f}_{gtr} =  f_{gtr}(\tilde{x}_{max}) = \frac{\kappa_1 \tilde{x}_m e_{\tilde{q}}(-1/\theta)}{e_{\theta}(-1/\tilde{q})} \; . 
\end{equation}
at 
\begin{equation}
\tilde{x}_{max} = \frac{1}{(\theta - \tilde{q})^{1/\theta}} = \frac{\tilde{x}_m}{e_{\theta}(-1/\tilde{q})} \; ,
\label{eq:pontomaximo}
\end{equation}
The control parameter $\kappa_1$ is unbounded ($\kappa_1 > 0$), for $\tilde{q} \ge 0$, but 
for $\tilde{q} < 0$, since $\tilde{f}_{gtr} \le \tilde{x}_m$,  it belongs to the interval $0 < \kappa_1 \le \kappa_{m}$, where:
\begin{equation}
\kappa_m = \frac{e_{\theta}(-1/\tilde{q})}{e_{\tilde{q}}(-1/\theta)} = e_{\theta}(-1/\tilde{q}) e_{ - \tilde{q}}(1/\theta) \; .
\label{eq:km}
\end{equation}

From Eq.~(\ref{eq:pontomaximo}), one sees that for $\theta < \tilde{q}$, $f_{gtr}(\tilde{x})$ does not have a hump, it is simply a monotonically increasing function of $\tilde{x}$, which characterizes the contest  models. 
At the critical $\theta = \tilde{q}$ value, the function $f_{gtr}(\tilde{x})$ starts to have a maximum value at infinity. 
For $\theta > \tilde{q}$, the function $f_{gtr}(\tilde{x})$ has a hump, with maximum value at $x_{max}$ [Eq.~(\ref{eq:pontomaximo})] such that as $\theta \rightarrow \infty$ then $x_{max} \rightarrow 1$.
 
\begin{figure}[ht]
\begin{center}
\includegraphics[angle=-90, width =.7 \columnwidth]{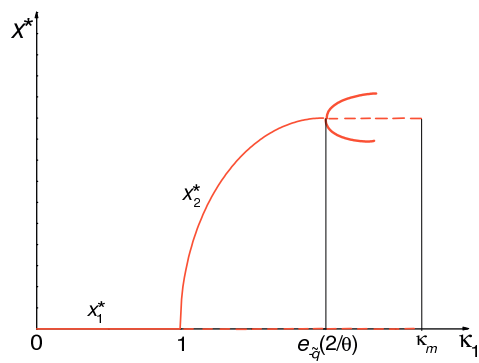}
\caption{Fixed points, given by Eqs.~(\ref{eq:x1}) and ~(\ref{eq:x2}), as function of $k_1$. 
Instability regions are represented by dashed lines. The value of $\kappa_m$ is given by Eq.~(\ref{eq:km}).}
\label{fig:cp}
\end{center}
\end{figure}

The map fixed points [$\tilde{x}^{*} = f_{gtr}(\tilde{x}^*)$] are:
\begin{eqnarray}
\tilde{x}_1^{*} & = & 0 \label{eq:x1}\\
\tilde{x}_2^{*} & = & [\ln_{\tilde{q}}(\kappa_1)]^{1/\theta} \ge 0 \; . 
\label{eq:x2}
\end{eqnarray}
These fixed points are show as function of $\kappa_1$ in Fig.~\ref{fig:cp}.

\begin{figure}[ht]
\begin{center}
\includegraphics[angle=-90, width = .7\columnwidth]{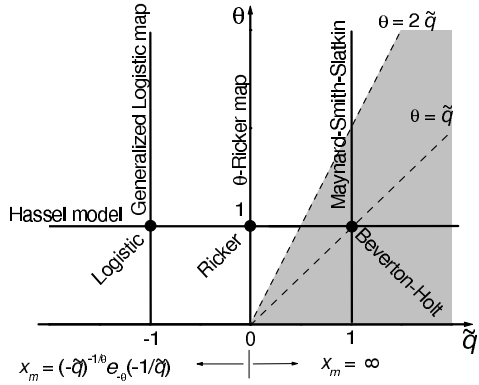}
\caption{Diagram distinguishing the several types of behavior of the generalized $\theta$-Ricker model [Eq.~(\ref{eq:final})]. 
For $\tilde{q} < 0$, one has the map $\tilde{x}_{i+1} = \kappa_1  \tilde{x}_i (1 - |\tilde{q}| \tilde{x}_i^{\theta})^{1/|\tilde{q}|}$ and $0 \le  \tilde{x} \le \tilde{x}_m$, with $x_m$ given by Eq.~(\ref{eq:xm}).
This system represents scramble models and achieves chaos, period doubling. 
For $\tilde{q} \ge 0$, $\tilde{x} \ge 0$ and the behavior of the system is driving by the $\theta$ values. 
For $\theta \ge 2 \tilde{q}$, we do no have analytical results. 
Numerical simulation indicate regions with finite order of period doubling. 
For $\theta < 2 \tilde{q}$, the system represent context models, with only two fixed points given by Eqs.~(\ref{eq:x1}) and~(\ref{eq:x2}). 
The map presents a hump, for $\theta >2\tilde{q}$, but for $\theta \le \tilde{q}$, the map is a monotonically increasing function. 
}
\label{fig:diagram}
\end{center}
\end{figure}

The fixed point $\tilde{x}_1^{*}$ represents the species extinction and is stable, for $0 < \kappa_1 < 1$.  
Both fixed points $\tilde{x}_1^{*}$ and $\tilde{x}_2^{*}$ are marginal, for $\kappa_1 =1$. 
For $1 < \kappa_1 < e_{-\tilde{q}}(2/\theta)$, $\tilde{x}_1^{*}$ becomes unstable and $\tilde{x}_2^{*} > 0$ is stable and represents the species survival.  
For $\kappa_1 = e_{-\tilde{q}}(2/\theta)$, $\tilde{x}_2^{*}$ becomes unstable and as $\kappa_1$ increases, a stable cycle-2 appears. 
For $\tilde{q} < 0$, as $\kappa_1$ increases further, the cycle-2 becomes unstable at some value of $\kappa_1$ giving rise to a route to chaos as in the logistic map, via period doubling. 
Nevertheless, for $\tilde{q} \ge 0$, several scenarios may take place. 
Even though $\tilde{q} > 0$ and $\theta > \tilde{q}$, if $\tilde{q} < \theta <  2\tilde{q}$, the map $f_{gtr}$ has a hump, but it is not thin enough to produce periods greater than unity.
In this case, $f_{gtr}$ produces only the two fixed points $\tilde{x}_1^{*}$ and $\tilde{x}_2^{*}$, which characterize the context models.   
Nevertheless, scramble models ($\theta > 2\tilde{q}$) have maps with a hump thin enough to produce stable cycles with period greater than unity. 
Thus, in scramble models one has  more complex scenarios such as period doubling, as a route to chaos, as $\kappa_1 > e_{-\tilde{q}}(2/\theta)$.  
We have not being able to obtain analytically the behavior of the system $\tilde{q} > 0$ and $\theta > 2 \tilde{q}$. 
The $\theta \times \tilde{q}$ diagram is depicted in Fig.~\ref{fig:diagram}.

For $\tilde{q} = \theta = 1$, one retrieves the Beverton-Holt model, with the fixed points $\tilde{x}_1^{*} = 0$ and $\tilde{x}_1^{*} = \ln_{1}(\kappa_1) = \kappa_1 - 1$. 
For $\tilde{q}$, one retrieves the generalized logistic map.

\section{Generalized Skellam model}
\label{sec:generalizedskellam}

All the contest competition models generalized by Eq.~(\ref{eq:generalized_theta_ricker_model}) are power-law-like models for $\tilde q \neq 0$. However, the Skellam contest model cannot be obtained from this approach. It is the complement of an exponential decay $x_{i+1}=\kappa(1-e^{-rx_i})$ \cite{skellam}. Nevertheless, it is interesting to replace the exponential function to the $\tilde q$-exponential in this model: $x_{i+1} = k [1 - e_{-\tilde q}(-rx_i)]$ and write $\tilde{x} = r x$ and $\kappa = r k$, which leads to:
\begin{equation}
 \tilde{x}_{i+1} = \kappa \left[1 - e_{-\tilde q}(- \tilde{x}_i)\right]\;.\
\label{eq:genskellan}
\end{equation}
For $\tilde q \rightarrow -\infty$, Eq.~(\ref{eq:genskellan}) leads to the constant model, for $\tilde q = -1$, the trivial linear growth is found. If $\tilde q=0$, one recovers the Skellam model and finally, $\tilde q = 1$ leads to the Beverton-Holt contest model (see Table \ref{tabela2}).

\begin{table}[hbt]
\begin{center}
\begin{tabular}{l|c}
\hline
\hline
Model                & $\tilde q$  \\
\hline
\hline
constant             & $-\infty$\\
\hline
linear               &  $-1$ \\
\hline
Skellam              & $0$\\
\hline
Beverton-Holt         & $1$\\
\hline 
\hline
\end{tabular}
\caption{Summary of the parameters to obtain contest competition discrete growth models from Eq.~(\ref{eq:genskellan}).}
\label{tabela2}
\end{center}
\end{table}

\section{Conclusion}
\label{sec:conclusion}

We have shown that the $\tilde{q}$-generalization of the exponential function is suitable to describe discrete growth models.
The $\tilde q$ parameter is related to the range of a repulsive potential and the dimensionality of the fractal underlying structure.
From the discretization of the Richard's model, we have obtained a generalization for the logistic map and briefly studied its properties.
An interesting generalization is the one of $\theta$-Ricker model, which allows to have several scramble or contest competition discrete growth models as particular cases.
Equation (\ref{eq:final}) allows the use of softwares to fit data to find the most suitable known model throughout the optimum choice of $\tilde q$ and $\theta$.
Furthermore, one can also generalize the Skellam contest model.
Only a few specific models mentioned in Ref. \cite{brannstrom_2005} are not retrieved from our generalization. 
Actually, we propose a general procedure where we do not necessarily need to be tied to a specific model, since one can have arbitrary values of $\tilde q$ and $\theta$.

\section*{Acknowledgments}
The authors thank C. A. S. Ter\c{c}ariol for fruitful discussions. 
ASM acknowledges the Brazilian agency CNPq (303990/2007-4 and 476862/2007-8) or support. 
RSG also acknowledges CNPq (140420/2007-0) for support. 
ALE acknowledges CNPq for the fellowship and FAPESP and MCT/CNPq Fundo Setorial de Infra-Estrutura (2006/60333-0).

\end{document}